\documentclass[12pt]{article}
\usepackage{amssymb}
\usepackage{amsmath}    
\usepackage{graphicx}
\usepackage{xcolor}
\usepackage{wasysym}

\title{The W boson mass: precision measurement and impact on physics}
\author{
Ashutosh V. Kotwal\\ 
\emph{Duke University} \\
}
\begin{document}
\maketitle
\begin{abstract}
As a mediator of the weak nuclear force, the $W$ boson influences many properties of fundamental particles and their interactions. 
 Understanding the $W$ boson as accurately as possible, including knowing its mass, has been a priority for decades. In a succession
 of measurements with increasing precision by multiple experiments, a significant tension between its measured and predicted mass has 
 recently been documented by the CDF Collaboration. 
 Smaller differences between measurements exist. As the $W$ boson mass provides a window on  new physics,
 a comparison between different measurement techniques can inform the path to further investigations. 
\end{abstract}
\section{Introduction}
\label{sec:intro}
\hspace*{0.18in}
Ever since the discovery of the weak nuclear force in radioactive decays more than 100 years ago, this interaction has continued to be 
a source of intrigue and new insights into fundamental particles and their interactions at the smallest distances. Among other 
 surprises, the weak nuclear force taught us that nature distinguishes between particles and their mirror images (so-called 
 parity violation) (see~\cite{wuExperiment} and references therein), and that fundamental particle masses are 
 emergent properties arising from their interaction with the constant,  
 non-zero value of the Higgs field in the vacuum~\cite{anderson,englert,pwhiggs,pwhiggs2,guralnik,atlasHiggs,cmsHiggs}. 

 The $W$ boson, as one of the carriers of the weak nuclear force, plays a special role in the well-developed theory of fundamental 
 particles referred to as the Standard Model (SM)~\cite{glashow,salam,weinberg}. Despite certain conceptual 
 limitations~\cite{gavinPerspective}, the SM continues to be extraordinarily 
 successful in predicting the properties of fundamental particles and their interactions as witnessed by the spate of results obtained 
 from experiments at the Large Hadron Collider (LHC)~\cite{atlas10Higgs,cms10Higgs}. 

 However, the recent, high-precision measurement of the mass of the $W$ boson~\cite{CDF2022} differs significantly $(7 \sigma)$ from its calculated 
 value in the SM~\cite{pdg,Ayres,Jens}, becoming the latest surprise in fundamental physics that points towards inadequacies of the SM 
 and opening 
 doors for further revelations. This measurement confirms the trend that almost all previous measurements have also been higher 
 than the SM expectation 
 (see Fig.~\ref{barPlot}); this measurement is the first to achieve the requisite precision to create a sharp inconsistency with the SM. 

The direct observation of the $W$~\cite{ua1Wdiscovery,ua2Wdiscovery} and $Z$~\cite{ua1Zdiscovery, ua2Zdiscovery} bosons, 
 the mediators of the weak interaction, at the CERN S$p \bar{p}$S collider
 in 1983 was a major milestone in the understanding of fundamental particles and their interactions. Since then, the precise
 measurement of their masses has been one of the priorities of the subfield of particle physics that focuses on precision
 measurements. In a few years following their discovery, their masses were measured with a precision of 
 2-0.5\%~\cite{ua1Wmass,ua2Wmass,ua2WmassFinal,CDFrun0PRL,CDFrun0PRD}. In the 
 following two decades, the large electron-positron (LEP) collider at CERN~\cite{alephWmass2000,delphiWmass2001,l3Wmass1999,opalWmass2001,ALEPH,DELPHI,L3,OPAL,lepwmass} and the Tevatron proton-antiproton collider 
 at Fermilab~\cite{CDFrun1aPRL,CDFrun1aPRD,CDFrun1bPRD,DZEROrun1aPRL,DZEROrun1aPRD,DZEROrun1bCCPRL,DZEROrun1bCCPRD,DZEROrun1bECPRL,DZEROrun1bECPRD,DZEROrun1cPRD} advanced these measurements. They surpassed 0.1\% precision in the $W$ boson mass~\cite{cdfd0Run1Combo} and achieved
 23 parts per million precision in the mass of the $Z$ boson~\cite{pdg}. With the precise measurements of the $Z$ boson mass, the lifetime
 of the muon and the coupling strength of the electromagnetic interaction, as well as the masses of the top quark and the 
 Higgs boson, the SM is able to calculate the 
 expected mass of the $W$ boson to an accuracy better than 0.01\%~\cite{pdg}. 

 Between 1995 and 2000, the Tevatron collider was 
 upgraded to deliver a factor of 100 more data than its 1992-1995 run. This second run operated between 2000 and 2011. Along 
 with many other physics goals, improving the precision of the $W$ boson mass measurement to test the SM was a priority for the Tevatron
 experiments CDF and D0. Using the steadily increasing datasets, both CDF~\cite{CDF2firstPRL,CDF2firstPRD,cdf2fbprl,cdf2fbPRD} 
 and D\O~\cite{dzero1fbprl,dzero5fbprl,dzero5fbPRD} produced multiple measurements 
 of the $W$ boson mass with increasing precision, reaching 0.02\% in 2012~\cite{run2combo}. 
 In parallel, the LHC started data-collection in 2009 and some of these data have been analyzed to publish $M_W$ measurements by the 
 ATLAS~\cite{atlasWmass,atlasWmassErratum} and LHCb~\cite{lhcb} experiments, achieving precisions of 0.02\% and 0.04\% respectively.   
 The latest result from CDF in 2022 uses the complete Tevatron dataset and achieves a precision of 
 almost 0.01\%~\cite{CDF2022}. This measurement reveals the $7\sigma$ tension with the
 SM expectation, prompting a flood of speculations on the implications for new physics. 

 The measurement of the $W$ boson mass is substantially more complicated than the $Z$ boson mass, even though both particles
 have clean two-body decay modes containing electrons or muons. Most of these complications arise from the presence of the 
 neutrino in the $W$ boson decay, which is undetectable while carrying away half of the rest-mass energy of the boson. 
 The inference of the latter from the partial information available in the final state is one of the reasons that each iteration
 of the data analysis has required many years\footnote{$W$ bosons also decay to quark-antiquark pairs. This mode has never been
 used to make a precise measurement of $M_W$ at a hadron collider, because this mode is swamped by non-$W$ background processes and 
 because the measurement of the net 4-momentum of the daughter particles is not sufficiently precise or accurate.}. 

 This article, which is part review and part perspective, recaps the role of $m_W$ in the SM and its extensions, 
 compares and contrasts $m_W$ measurement techniques, and discusses prospects and future directions.  

\section{Theoretical motivation}
\label{sec:motivation}
\hspace*{0.18in}
 The mass of the $W$ boson is a very interesting observable because there is a strong theoretical connection between 
 the masses of the $W$ and $Z$ bosons in the SM, a relativistic quantum field theory of matter and its electromagnetic, weak and strong  
 interactions. To understand this connection, we can imagine an inaccurate but simplest version of the weak
 interaction based purely on the SU(2) gauge group. In the SM, the principle of gauge symmetry of 
 the fermionic wave functions is invoked, under which the fermions undergo transformations in an internal space while maintaining 
 the invariance of the theory. The transformation 
 is referred to as a gauge transformation when it is a smooth (but otherwise arbitrary) function of spacetime. The manifold of
 these symmetry transformations is described by a chosen Lie group, which is referred to as the gauge group. 

 The simplest gauge group is U(1) whose manifold is a circle; therefore this gauge transformation changes the phase of the wave function 
 smoothly with respect to spacetime. The principle of gauge symmetry states that the fermion's equation of motion should
 be invariant under these arbitrary transformations. This is accomplished by corresponding changes in the induced vector
 potential by the spacetime derivative of the arbitrary phase function. The theory with this choice of gauge group describes fermions
 with electromagnetic interactions. The generators of the Lie group are identified with the vector bosons that mediate the interaction;
 the U(1) generator is identified with the photon. 

 When the gauge group is SU(2) (whose manifold is the unit 3-sphere $S^3$), the fermions are described in the internal space by two 
 components that transform into each other under the action of the 
 symmetry. As a gauge transformation, it is again local, i.e. it is a matrix-valued function of spacetime. Therefore, the induced 
 vector potential is 
 also matrix-valued, and couples the two components of the fermion wave function to each other. For the weak interaction, the vector
 bosons thus mediate transformations between the electron and the electron-neutrino, between the up and down quarks and in general 
 between the two components of the weak-SU(2) doublet of fermions.  

 Infinitesimal SU(2) group transformations have a correspondence with infinitesimal rotations of a sphere. The basis vectors of 
 a two-component fermionic wave function may be visualized as the poles of the sphere\footnote{This analogy captures some of the 
 properties of the SU(2) action, but not all of them.  Precise mathematical discussions can be found elsewhere.}. 
 This is in analogy with SU(2) being used to describe rotations of electrons in space, where the  components of the 
 spinors of SU(2) describe the spin-up and spin-down states of electrons. 
 Continuing this analogy, the weak interaction transforms a fermion doublet from one component to the other by the emission or absorption 
 of a $W$ boson.  This description requires the existence of a third transformation that can be visualized as a rotation
 along the equator of the sphere. Thus, the SU(2) gauge group predicts three gauge bosons, which mathematically serve as the generators
 of group transformations and physically mediate all possible weak interactions. 
 In this simplest picture, the three types of interactions associated with the weak force must be exactly symmetric because
 they are related by the rotational symmetry. 

 The principle of symmetry under gauge transformations (called gauge invariance) 
 predicts that these transformations must be induced by 
 massless gauge bosons in the relativistic quantum theory. This principle has served to beautifully explain the massless
 photon as the gauge mediator of the electromagnetic interaction, and the massless gluons as the mediators of the 
 strong nuclear force when probed at high energies. 

 The weak interaction, on the contrary, appears to be mediated by very massive bosons, which explains why this interaction
 appears to be ``weak'' at low energies. This mystery led to the postulate of the Anderson-Brout-Englert-Higgs-Guralnik-Hagan-Kibble 
 mechanism~\cite{anderson,englert,pwhiggs,guralnik} in which the Lorentz-invariant Higgs field develops a constant, non-zero value in 
 the vacuum. The interactions of all particles 
 experiencing the weak force with this Higgs condensate imparts the emergent property of mass to them as a low-energy
 manifestation. 

 In this simplest description of the weak-interaction group, the three gauge bosons will acquire the same mass because 
 of their mutual rotational symmetry - the two  $W$ bosons and the third would-be $Z$ boson. From this starting point, 
 we can sequentially incorporate the known principles of nature that modify the original rotational symmetry. The biggest
 effect is the quantum-mechanical mixing between the third SU(2) generator and another, electromagnetism-like U(1) interaction called 
 hypercharge, yielding 
 the observed $Z$ boson and the photon ($\gamma$) as the mixed states~\cite{glashow}. The Higgs field is an SU(2) doublet. The  
 quantum numbers of its condensate are such that one of the mixed states  
 remains massless; this state is identified as the photon~\cite{salam,weinberg}. In other words, while the condensate (i.e. the true vacuum) is charged
 under SU(2)$_{\rm weak}$ and under U(1)$_{\rm hypercharge}$, the linear combination of these charges that is identified as electric
 charge is zero for the condensate. Therefore the vacuum is electrically neutral, gauge invariance is maintained for electromagnetism, 
 electric current is conserved and the photon is massless. 
 However, the weak charge and hypercharge of the Higgs condensate ensure that the $W$ and $Z$ bosons acquire masses  whose 
 values are proportional to  
 their respective dimensionless coupling to the condensate and the energy-dimensioned value of the latter. This means that the 
 weak current and the hypercharge current are not separately conserved after the Higgs condensate forms~\cite{salam,weinberg}. 

The electromagnetic fine structure coupling $\alpha$, the Fermi coupling constant $G_F$ 
 (extracted from lifetime of the muon) and the $Z$ boson mass are the most precisely measured observables~\cite{pdg} related to the three 
 parameters of the theory; the SU(2) and U(1) gauge couplings and the value of the Higgs condensate. With the tight constraints on the
 parameters from these measurements, the theory is rendered predictive for $M_W$. In other words, the theory predicts $M_W$ 
 when $\alpha$, $G_F$ and $M_Z$ are fixed to their measured values~\cite{marcianoSirlin}, 
\begin{equation}
M_W^2 (1 - \frac{M_W^2}{M_Z^2}) = \frac{\pi \alpha}{\sqrt{2} G_F} (1 + \Delta r) \nonumber
\end{equation}   
 where $\Delta r$ denotes the quantum-mechanical radiative corrections (explained in the next paragraph) and
  classical physics predicts $\Delta r = 0$. 
 The mixing between one of the SU(2) generators and the U(1) generator violates the original rotational 
 symmetry (called custodial symmetry) between the three SU(2) generators; the $Z$ boson mass differs from the $W$ boson
 mass, with $M_Z > M_W$ since $M_\gamma = 0$.  $M_W$ is thus a probe of such custodial symmetry-violating effects, which makes its 
 precise measurement of high importance to check for a deviation from the calculated value~\cite{rossVeltman}.

The  SU(2)$\times$U(1) mixing effect is within the realm of classical physics. In the 1990's, the 
 precision of the $M_W$ measurement became sufficient to discern the small quantum effects denoted by $\Delta r$; these are due to 
 vacuum fluctuations wherein particle-antiparticle pairs
 appear and disappear over very short time intervals, consistent with the Heisenberg uncertainty 
 principle and special relativity. The gauge bosons experience quantum fluctuations into fermion-antifermion pairs~\cite{hooftVeltman}. 

 From the perspective of the $W$ boson mass,  the 
 most interesting fluctuations involve the top quark and the bottom quark, members of a weak-SU(2) doublet. However, their 
 interaction with the Higgs field is very different, resulting in a large difference in their masses and also a large violation of the 
 custodial symmetry that tries to maintain the similarity of the $W$ and $Z$ boson masses.  Hence, quantum fluctuations in the gauge 
 boson propagators involving weak-SU(2) doublet components with different masses   
 induce a calculable modification of $M_W$ for a fixed $M_Z$; this modification is dominated by fluctuations into top-bottom quarks 
 which have by far the largest mass difference among all SU(2) doublets. In other words, mass differences between weak-SU(2) doublet 
 fermions represent another violation of the custodial symmetry and therefore induce a shift in $M_W$ through quantum 
 effects~\cite{veltmanRho}. 

Historically, an estimate of the top quark mass was inferred from such quantum corrections~\cite{langacker}. Since
 the top quark discovery by the CDF~\cite{cdfTopDiscovery} and D\O~\cite{d0TopDiscovery} experiments at the Tevatron in 1995, its mass has been measured directly and included as an input in the 
 calculation~\cite{pdg}. Finally, quantum corrections involving the Higgs boson also induce a shift in the expected value of 
 $M_W$~\cite{veltmanHiggsLoops}, 
 and this became a method for inferring the Higgs boson mass before the latter was directly observed. 

 This history illustrates how the $W$ boson mass, in conjunction with other measurements, has
 been a harbinger of fundamental principles of nature. Looking to the future: with increasing precision, $M_W$ can probe additional laws of physics that may follow this pattern of custodial-symmetry violation
 between the $W$ and $Z$ bosons. Hypotheses include a new, weak force (which might mediate an interaction with 
 dark matter) that mixes with the photon and the $Z$ boson, the existence of additional Higgs-like bosons that interact
 differently with $W$ versus $Z$ bosons, and heavy, weakly-interacting fermion or boson pairs that have a large mass difference 
 akin to the top-bottom mass difference. 

 A popular hypothesis that accommodates the latter is supersymmetry, in which all SM 
 particles have supersymmetric partners whose intrinsic spin differs by $\hbar/2$ from their respective SM counterparts\footnote{ The reduced  Planck's constant $\hbar \equiv h/(2\pi)$ is the fundamental unit of angular momentum.}. 
 Such heavy particles may also arise as composite states due to substructure and a new strongly confining interaction at the TeV energy scale. This new interaction and substructure may provide a dynamical explanation for the formation of the Higgs condensate,  
 and the Higgs boson may be a bound state rather than the fundamental scalar of the SM. 
 Conceivably, this new dynamical sector is connected with 
 dark matter particles, or with the mystery of the excess of matter over antimatter in the universe, or both. Both supersymmetry and 
 compositeness allow for a new conserved quantum number that stabilizes one or more of the new particles which, if interacting very 
 weakly with SM particles, can constitute the dark matter.         

 These are some of the hypotheses being investigated as explanations of the difference between the CDF measurement and the SM 
 expectation of $M_W$. 
\section{Measurement of the $W$ boson mass}
\label{sec:mass}
\hspace*{0.18in}
 The $W$ boson mass has been measured both at proton-(anti)proton colliders and electron-positron 
colliders. It is healthy for the scientific development of the field to perform measurements under different 
 experimental circumstances and with different techniques. 

\begin{figure}[htbp]
\begin{center}
\vspace*{-145mm}
\includegraphics[width=17cm]{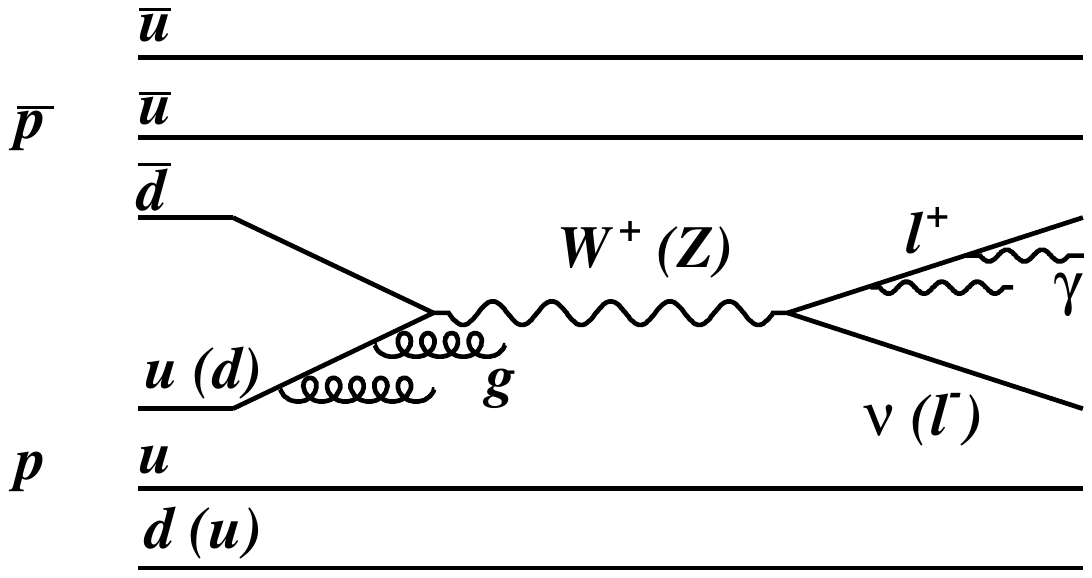}
\caption{Illustration (reproduced with permission from Ref.~\cite{cdf2fbPRD}) of $W$ (or $Z$) boson production due to quark ($q$) -- antiquark ($\bar{q}$)  
 annihilation in $p\bar{p}$ collisions at the Tevatron, where the initial-state $q$ ($\bar{q}$) are predominantly intrinsic to 
 the proton $p$ (antiproton $\bar{p}$). Quarks associated with the annihilation to a $Z$ boson are indicated in parentheses.
  Initial-state gluon ($g$) radiation and  final-state photon ($\gamma$) radiation are also shown.
 The initial state QCD radiation fragments into a spray of hadrons. 
 At the LHC, the initiating $\bar{q}$ arises from $g \to q \bar{q}$ splitting. 
}
\label{fig:wzprod}
\end{center}
\end{figure}

The largest samples of $W$ and $Z$ bosons are produced at proton-(anti)proton colliders. The bosons are mostly produced
 singly from the annihilation of a quark and an antiquark respectively from the colliding beam particles (see Fig.~\ref{fig:wzprod}). 
 These initial-state particles are strongly-interacting, i.e. the principle of gauge symmetry with the gauge group SU(3) explains the 
 ``strong'' interactions between quarks as being mediated by gluons. Analogous to the electric charge associated with particles that
 interact electromagnetically, the ``charge'' quantum number associated with the strong interaction is called color. When a 
 colored quark and antiquark collide to produce a colorless
  electroweak boson, they are decelerated which causes them to radiate gluons. Thus, 
 electroweak gauge boson production is accompanied by multiple radiated gluons which further split into quark-antiquark pairs. This 
 shower of colored particles coalesces into a spray of color-neutral bound states called hadrons due to the color-confining nature
 of the strong interaction. 

 The decay products of the $W$ boson provide insufficient information to fully
 reconstruct the quark-antiquark collision, due to the invisible momentum carried away by the neutrino. The measurement of the net momentum
 carried by the accompanying spray of hadrons is therefore necessary to recover as much of the missing 
 information as possible. 

The SU(3)-gauge theory of strong interactions called quantum chromodynamics (QCD) has made enormous progress over the last five decades.
   In one of its great insights, the combined effect of quantum fluctuations involving quarks and gluons is to reduce
 the effective color charge as the energy of the scattering process increases~\cite{grossWilczek,politzer}. 
 This property of asymptotic freedom of 
 QCD makes it possible to calculate high-energy QCD-radiative processes in perturbation theory. However, perturbation theory becomes
 progressively inaccurate when the relevant process energy decreases to the GeV level and the effective QCD coupling is no longer small
 enough. 

\subsection{Sensitivity to QCD calculations}
\label{sec:qcd}
\hspace*{0.18in}
The inference of the neutrino momentum 
 is aided by the use of theoretical calculations of $W$ boson production and decay, that take into 
 account the effects of the strongly-interacting QCD radiation. Over the decades, these calculations have progressed in accuracy
 by including quantum-mechanical amplitudes at higher order in perturbation theory. 

 An important theme in the $M_W$ measurement 
 is the ability to improve the accuracy of these calculations by applying constraints from the data. In this respect,
 the Tevatron has some advantages over the Large Hadron Collider (LHC).  
 At the Tevatron, antiprotons already contain antiquarks, whereas the latter have to be generated through the strong interaction
 radiative processes when using proton beams at the LHC~\cite{atlasWmass,atlasWmassErratum}. Second, the fractional 
 momenta of the (anti)quarks with respect to the (anti)protons at the Tevatron are larger, and their probability distributions are better 
 measured and less affected by 
 QCD radiation, in comparison to the corresponding distributions at the LHC~\cite{CDF2022}.  
 Third, $W$ boson production at the Tevatron is minimally induced by 
 the heavy bottom, charm and strange quarks, which do contribute significantly at the LHC~\cite{atlasWmass,atlasWmassErratum} and whose effects are difficult to calculate. 

 Within the general theme of constraining the QCD modeling of $W$ boson production with measurements of the accompanying 
 QCD radiation, there are differences between the methodologies used by the Tevatron experiments (CDF and D0) and the 
 LHC experiments (ATLAS and LHCb, with the first measurement from the CMS experiment to be expected). 

 The Tevatron experiments have consistently used a calculation of vector boson production that includes the effects 
 of multiple soft-gluon radiation and gluon splitting into quark-antiquark pairs, coupled to an exact calculation of 
 a single energetic gluon emission or gluon splitting~\cite{resbosRef1,resbosRef2,resbosRef3}. There has been discussion of this boson production and decay 
 calculation in the context of the recent, highest-precision measurement of the W boson mass by the CDF experiment~\cite{CDF2022,resbos2}. One
 of the discussion points is the effect of double gluon radiation or gluon splitting with high transverse momentum ($p_T$).
  Three facts shed light on this point. 

 First, the phase space of the $W$ bosons selected for this analysis is confined to 
 small $p_T$, where the ``resummation'' of multiple gluon emission and splitting is the most relevant 
 description of the boson production process. The so-called ``fixed-order'' calculation of individual emissions is already
 a minor correction in this phase space. Second, the calculation is tuned on the observed $p_T$ spectrum 
 of the $Z$ boson, which constrains the QCD parameters that are largely identical between $Z$ and  $W$ boson
 production. All hadron collider experiments follow this strategy.

 Third, a unique feature of the CDF measurement is the direct application of the $W$ boson $p_T$ data as an 
 additional constraint on the corresponding calculation. The perturbative QCD 
 calculation of the $W$ boson $p_T$ spectrum may not be sufficiently accurate even after the calculation has been adjusted so that 
 the calculated and the measured $Z$ boson $p_T$ spectra agree. The residual variation in the calculated $p_T^W$ spectrum due to 
 unconstrained degrees of freedom has an impact on the $M_W$ measurement. However, no other experimental analysis has constrained 
 the calculation of the $W$ boson $p_T$ spectrum with the corresponding observed distribution in the signal sample, as CDF has. 

This constraint would be difficult to apply for the LHC
 experiments because of the degraded resolution on $p_T^W$ due to the large number of
 additional beam-beam  interactions occurring simultaneously (called pile-up events) 
 at the LHC. Tevatron experiments have the advantage of running
 with lower pile-up, which permits CDF to measure $p_T^W$ with adequate resolution to apply this constraint. 

 In this context, the details of $W$ versus $Z$ boson production merit a discussion. For massless initial-state quarks 
 (a valid approximation at the Tevatron), CDF has used the {\sc dy}{\small q}{\sc t} calculation~\cite{dyqt1,dyqt2} to confirm that the $p_T^W$ and $p_T^Z$ spectra 
 would be identical if the bosons had the same mass. Therefore, the difference between $W$ and $Z$ bosons
 in the flavor composition of the initial-state quarks does not impact perturbative-QCD 
  processes. The difference in the boson mass scales, however, does change the phase space available for QCD radiation and 
 impacts the $p_T^W/p_T^Z$ ratio of spectra. The variation in this ratio  
 can be parameterized in terms of the renormalization, factorization and resummation 
 scales (in increasing order of sensitivity) which capture the perturbative inaccuracy of the QCD calculation. 
 CDF constrained this parameterization using the measured $W$ boson $p_T$ spectrum. In 
 the context of perturbative QCD, the concern that the tuned $Z$ boson production 
 model may not be portable to $W$ bosons has been addressed directly only by the CDF analysis. 

In the ATLAS analysis~\cite{atlasWmass,atlasWmassErratum} an attempt is made to check the theoretical $p_T^W/p_T^Z$ spectrum ratio against a low-precision
 measurement made with $\cal O$(1\%) of the $M_W$-measurement dataset. Calculations are also compared with the tail of the 
 observed distribution of the component of $\vec{p}_T^{~W}$ along the charged-lepton direction. This quantity is presented as a kind of 
 proxy for $p_T^W$ and used to reject certain calculations of the $p_T^W/p_T^Z$ spectrum ratio.  

In the LHCb analysis~\cite{lhcb}, a measurement of $\vec{p}_T^{~W}$ is not possible because the detector does not surround the beam collision
 point. An attempt is made to gain sensitivity to the $p_T^W$ spectrum by floating an additional nuisance
 parameter in the fit to the $q^{\mu}/p_T^{\mu}$ spectrum of muons (of charge $q^{\mu}$) from $W^\pm \to \mu^\pm \nu$ data. This parameter converges to a different
 value from the one that best fits the $p_T^Z$ data. If they are constrained to a common value, the inferred $M_W$ increases by 39~MeV, 
 though the quality of the fit degrades. This situation illustrates why a direct constraint from $p_T^W$ data, as obtained by CDF, 
  is preferable to the use of a proxy or multivariate fits with hidden dependencies that are difficult to control.  

 At very low boson $p_T$, the effective strong-interaction coupling becomes large enough that non-perturbative QCD processes 
 dominate, which may differentiate between $Z$ and $W$ bosons. In the limit of massless quarks, this possibility is 
 considered in the CDF analysis by including a mass-dependence in the non-perturbative parameters. One parameter is constrained 
 with the CDF data on the $Z$ boson, and a published global fit is used for 
 the second parameter. The global fit includes the measured $p_T$ spectra of low-mass vector bosons~\cite{resbosRef3}. 
 As this global fit was performed about 20 years ago, it would be useful to repeat it with new data on low-mass boson production and 
 with newer calculational tools, to help confirm the robustness of 
 the CDF analysis in this respect. 

 In summary, all hadron collider analyses of $M_W$ tune parameters of the QCD calculation of boson production to match the observed
 $p_T^Z$ distribution. Other than CDF, all other analyses assume that the tuned calculation automatically describes $W$ boson production
 accurately, though additional parameters are known to differentiate between $W$ and $Z$ boson production. What sets the CDF analysis 
 apart is its use of the observed $p_T^W$ distribution to constrain the most important of these parameters, demonstrating its reliability
 in this respect.  
\subsubsection{Theoretical issues in non-perturbative QCD}
\label{sec:nonPertQCD}
\hspace*{0.18in}
 The preceding discussion of electroweak boson production, framed in the context of massless quarks, suggests the following
  investigation. 
 A  fraction of $W$ bosons are produced with a charm quark in the initial state, whose 
 mass of $\approx$1.3~GeV may induce a non-perturbative effect that differentiates $W$ from $Z$ boson production and spoil
 the portability of the $Z$ boson production model to $W$ bosons. At the 
 Tevatron, only a small percentage of $W$ boson production is induced by charm quarks, while this fraction is many times larger
 at the LHC. The effect of the charm quark mass is expected to be negligible at the Tevatron; a definitive confirmation
 requires a QCD calculation that includes the charm quark mass. Such a calculation would be a welcome addition in the domain 
 of high-accuracy QCD calculations of electroweak boson production. Similar considerations apply for bottom quark-induced $Z$ 
 boson production at the LHC. Since heavy-quark mass effects are
 a source of systematic uncertainty at the LHC, $M_W$ measurements at the LHC would benefit. 
  
 Another aspect of non-perturbative QCD is worth exploring. The wave functions of up and down quarks in nucleons 
 are governed by non-perturbative QCD dynamics. Since there is a net of two up quarks and one down quark in the proton, 
  Fermi-Dirac statistics suggests a difference between the up- and down-quark wave functions and their intrinsic $p_T$. 
 As the mix of up and down quarks differs between $W$ and $Z$ bosons, the intrinsic $p_T$ difference may lead 
 to small differences in the $W$ and $Z$ boson $p_T$ spectra at small $p_T$~\cite{soper}. This is another example of 
 non-perturbative effects that can spoil the  portability of the $p_T^Z$ spectrum parameters to the 
 $p_T^W$ spectrum. The quark-flavor dependence of the intrinsic $p_T$ is not known; its theoretical investigation would benefit all 
 hadron-collider analyses of $M_W$. 

 Thus, improvements in the theoretical understanding of heavy quarks and of the flavor-dependent 
 non-perturbative aspects of proton structure would help to port the $Z$ boson model to $W$ bosons at very low $p_T$.   
\subsection{Momentum calibration of the experiment}
\label{sec:pcalib}
\hspace*{0.18in}
 One of the most important aspects of the $M_W$ analysis is the calibration of the momentum of the charged lepton 
 (electron and muon) emitted in the $W$ boson decay. Charged-particle momenta are measured using a magnetic tracker,  
 consisting of position-measuring sensors arranged in a concentric cylindrical configuration and immersed in an axial magnetic
 field. The position measurements along the helical trajectories are used to infer the curvature, the direction, and 
 in conjunction with the magnetic field, the momentum vector. Electrons and photons are measured using electromagnetic calorimeters 
 that convert the 
 particle's energy into a shower of soft photons and electron-positron pairs in the nuclear electric field of a dense material. As 
 this cloud of charge  traverses instrumented active material, it
  creates  scintillation light or ionization which is converted to a proportional electrical signal as a measure
 of the incident electron/photon energy.   

 The momentum calibration philosophies followed by the hadron-collider 
 experiments fall into two categories, one followed by the ATLAS, D\O\ and LHCb experiments and the other followed by the CDF 
 experiment. 

 The ATLAS and D\O\ experiments make no attempt to calibrate the magnetic tracker from first principles. Instead, they 
 calibrate solely on the $Z$ boson mass, i.e. adjust the detector response functions until the 
 $Z$ boson mass measured using the adjusted lepton momenta agrees with the previously measured precise value 
 from the LEP collider. At first glance this procedure seems sound since $M_Z = 91.19$~GeV is similar to $M_W \approx 80.4$~GeV and
 the calibration has to be ported over a fraction of the relevant momentum range. 

 However, the difference between $M_W$ and $M_Z$ is  11~GeV; when the $M_W$ measurement reaches the precision level of $10^{-4}$ there is no {\em a-priori} proof that porting the $M_Z$-based calibration to $M_W$ is fundamentally understood at the  $10^{-3}$ level. 
 In more than three decades of $M_W$ measurements at hadron colliders, the sole reliance on $M_Z$ for calibration has never been 
 formally proven to be sufficiently accurate. 

 The $M_W$ analysis from the complete Tevatron Run 2 dataset by the D\O\ experiment is an instance where the $M_Z$-based calibration
 strategy did not ultimately succeed. The D\O\ experiment has not analyzed the latter 40\% of Run 2 data, apparently
  due to tracker degradation~\cite{d0tracker}. The larger implication is that calibration on $M_Z$ alone is insufficient for the 
 requisite accuracy and robustness, due to a dependence on tracker information. This operational experience and its relevance for 
 $M_W$ measurement would be useful to document. 
 
 Cognizant of potential weaknesses of a solely $M_Z$-based calibration, all three CDF publications from Run 2 data emphasized 
 a more robust calibration and analysis strategy from the beginning. 
 In contrast with the other analyses, the CDF procedure for charged-lepton momentum calibration 
 relies on fundamental principles of operation of the sensor technologies. The first step is the alignment and 
 calibration of the drift chamber~\cite{cotNim}, which yields the calibrated track momentum of all charged particles. CDF has developed
 a special technique using cosmic rays~\cite{cotcosmic} to pin down the positions of the 30,240 sense wires in the drift chamber to a
 precision of $\approx 1 ~ \mu$m, 
 eliminating many biases in the measurement of track parameters~\cite{cosmicAlignment}. The procedure was refined over three iterations in the
 last two decades, each iteration requiring between one and two years. 

 Cosmic rays provide the most direct and transparent method of aligning the position-measuring sensors. As these muons
 rain down on the detectors from the atmosphere, they traverse the tracking sensors diametrically across their cylindrical geometry. 
 This trajectory builds in a powerful redundancy because the same muon is measured twice, on opposite sides of the beam axis. In the 
 CDF procedure, a single helical fit is performed to this double-sided trajectory~\cite{cotcosmic}. This fit is immune to many of the deformations
 of the tracker that bias collider tracks; the latter are prone to bias because they propagate outwards from the beam axis. Due to this 
 intrinsic rigidity, the 
 double-sided fitted trajectory of the cosmic ray provides the most robust reference with respect to which the sensors can be aligned, 
 thus eliminating the deformation-induced bias~\cite{cosmicAlignment}.
 Given the large size of the cosmic-ray sample at CDF, different sensors overlap sufficiently that they are all anchored with 
 respect to each other with a precision of $\approx 1 ~ \mu$m. 

 The CDF procedure emphasizes the importance of control; the observable(s)  chosen for calibrating any parameter(s) should (i) be a 
 strong function of the parameter(s) of interest, and (ii) be as independent as possible of other variables. The latter is a 
 crucial consideration because it minimizes the chance of being misled in the parameter inference. Both requirements are strongly
 satisfied by the CDF cosmic-ray alignment procedure -- it isolates the degrees of freedom of the sensor positions and is almost 
 independent of other variables. It is then possible for CDF to analytically identify the remaining modes of 
 displacement that cannot be measured by the cosmic-ray method; these are pinned down using complementary techniques~\cite{CDF2022}. These features
 distinguish the CDF tracker alignment from the procedure published by, for example, the LHCb experiment~\cite{lhcb}. 

 The next step in the CDF procedure is the calibration of the magnetic 
 field and its spatial non-uniformity using the reconstructed invariant mass of the $J/\psi$ particle, whose mass is precisely known,
 in its dimuonic decay channel. This large sample enables a fine-grained spatial scan of the drift chamber. 
 These muons also 
 cover a wide range in track curvature, the quantity directly measured
 by the drift chamber, and enable the calibration of momentum-dependent effects such as the ionization energy
 loss in the passive material~\cite{CDF2022,CDF2firstPRD,cdf2fbPRD}. 

 Additionally, CDF uses a large sample 
 of $\Upsilon$ particles decaying to dimuons to demonstrate the consistency and robustness of the tracker
 calibration. First, $\Upsilon$s with a higher (and again precisely known) mass than the $J/\psi$ cover
  the phase space closer to $W$ bosons. Second, the muon tracks used for the $\Upsilon \to \mu^+ \mu^-$ mass calculation can be
 reconstructed with two variants of the track reconstruction algorithm. The consistency of the measured $\Upsilon$ mass 
 between these variants is another important demonstration of the robustness of the momentum 
 calibration~\cite{CDF2022,CDF2firstPRD,cdf2fbPRD}. 

 Instead of a purely empirical calibration of the electromagnetic calorimeter, CDF performs a detailed first-principles simulation
 of electron and photon detection~\cite{calgeantnim} 
 based on the knowledge of the calorimeter geometry and electromagnetic showering calculations encoded in the {\sc geant} 
 program~\cite{GEANT1,GEANT2}. Next,
 CDF uses the electrons emanating from $W$ boson decay, with their calibrated momenta from the tracker, as an {\em in-situ} test-beam to
 measure the calorimeter thickness, non-linear energy response due to scintillator ageing, spatial and temporal variations in 
 response, and the  amount of passive radiative material upstream of the tracker. Almost all tuned parameters are causally related 
 to the principles governing electron and photon interactions with various materials~\cite{CDF2022,CDF2firstPRD,cdf2fbPRD}. 
 
The ultimate confirmation of these calibration procedures is the independent measurement of the $Z$ boson mass by CDF in both 
 electron and muon channels, using the same ``blinded'' procedure as used for the $M_W$ measurement, and demonstrating consistency 
 with the LEP value. Amongst all hadron-collider 
 measurements of the $W$ boson mass, the three CDF measurements published from the Tevatron Run 2 data are unique in 
 pursuing  
 these high-precision calibrations and cross-checks from first principles~\cite{CDF2022,CDF2firstPRD,cdf2fbPRD}. All other hadron collider measurements have 
 been based solely on rescaling various observables based on the reconstructed $Z$ boson mass.  

 The LHCb experiment uses a mix of $J/\psi \to \mu^+ \mu^-$, $\Upsilon  \to \mu^+ \mu^-$ and $Z  \to \mu^+ \mu^-$ data 
 to understand the tracker. Misalignments are parameterized by corrections to the track curvature and the parameters are fitted using 
 a proxy for the reconstructed $Z$ boson mass in the data. The momentum scale and resolution parameters are fit simultaneously to the 
 $J/\psi, \Upsilon$ and $Z$ boson data; however the mutual consistency of the momentum calibration from independent datasets is not
 demonstrated~\cite{lhcb}. 

For the $M_W$ measurement, 
 the CDF strategy benefits from unique features of its drift chamber geometry and construction, features that are lacking in 
 the trackers of the other hadron-collider experiments. The CDF drift chamber is a single gaseous volume with embedded sense wires. 
 Groups of twelve wires (cells) are mounted on precision-installed cards at each end of the cylindrical chamber, substantially
 reducing the number of alignment degrees of freedom~\cite{cotNim}. In comparison, the LHC experiments  use silicon sensor-based trackers 
 which consist of many tens of thousands of small modules, leading to an explosion in the alignment degrees of 
 freedom and a corresponding increase in tracker complexity for fundamental understanding. 

The second advantage conferred by the CDF drift chamber is that, as a charged particle traverses 96 sense wires radially, there are 
 no gaps or discontinuities in the instrumented volume. All particles must pass through 96 drift regions in which their positions 
are recorded, without exception. In particular, for incremental changes in the particle's curvature or direction, the relevant active 
 regions change continuously and smoothly, with a handful of transitions between adjacent drift cells. This is the reason the 
 curvature response function for this drift chamber is analytic, i.e. the measured track curvature is a smooth function of the 
true curvature and can be expressed as a polynomial function. Pinning down the few relevant polynomial coefficients is at the heart
 of the CDF tracker calibration. 

In comparison, silicon trackers provide a factor of ten 
  fewer (but substantially more precise) measurements, with many more boundaries between instrumented regions. It is more difficult to guarantee 
an analytic curvature response with few measurements along the track and a highly tiled detector geometry, which is inevitable when
 covering a large area with numerous small planar sensors.   

To be sure, silicon trackers provide considerably higher performance than drift chambers in terms of precision, rate capability, 
 granularity, radiation hardness and the ability to operate in high-flux environments. They are also uniquely suited to be placed close
 to the beam interaction point and measure the track impact parameters with high precision. It is the geometrical complexity and segmentation of silicon trackers that the curvature measurement is particularly sensitive to, because curvature is related to the second radial derivative  of position whereas impact parameter is related to the first derivative. And for the $m_W$ measurement, it is the track curvature that is 
 crucial for the accurate inference of the transverse momentum, while the impact parameter is irrelevant because the daughter leptons
 are prompt. 
  
 This discussion motivates further studies of tracker geometries and an exploration of geometrical deformations that lead to a 
 non-analytic curvature response  at small curvature. It behooves the experiments to conduct such studies in order to fully 
 understand and exploit magnetic trackers for precision measurements. It would be natural for CDF to lead these studies as they are 
 most advanced in the systematic understanding of tracker-based momentum calibration. 

 In the spirit of making the most reliable measurement possible, it is crucial that the $M_W$ analysis be based on fundamental 
 principles which are transparent and are in the public domain. CDF's strategy for calibrating the charged-lepton momentum exemplifies this 
 philosophy. 
\subsection{Event selection}
\label{sec:selection}
\hspace*{0.18in}
The upside of $W$ boson production at hadron colliders is that the production rate is substantially higher than at electron-positron 
 colliders (e.g. LEP II); the complication is that this rate is less than one-millionth of the total hadron-hadron 
 interaction rate. Real-time filtering to record events of interest while rejecting 99.99\% of the uninteresting collisions is a crucial
 aspect of the experimental techniques at hadron colliders. In the offline analysis of the recorded data, further requirements are 
 placed on the event information to suppress high-rate backgrounds from QCD-induced non-$W$ events that mimic high-$p_T$ electrons/muons  
 along with a $p_T$ imbalance due to resolution, faking a neutrino signature. 

It is important to minimize the selection bias on $W$-boson events while also minimizing the misidentification backgrounds. 
 Both effects alter the shape of the kinematic distributions from which $M_W$ is inferred. 
 The larger the bias, the more accurately it must be estimated from control samples of data
 or high-quality simulation. 

Of the hadron collider experiments, CDF has the highest charged-lepton trigger efficiency and the smallest misidentification backgrounds; 
 for $W$ boson events selected with the charged-lepton $p_T^{\ell} > 30$~GeV for the analysis, the trigger efficiency is essentially 100\% with 
 no dependence on $p_T^{\ell}$. As shown in Fig.~S40 of~\cite{CDF2022}, the inferred $M_W$ is stable with respect to the  
 $p_T^{\ell}$ requirement; no correction for trigger inefficiency is needed here. 
 This is a substantial simplification compared to the $p_T$-dependent trigger inefficiencies in the ATLAS, D\O\ and LHCb experiments, 
 which must be estimated and corrected for to prevent biasing the inferred $M_W$. The dependence of the selection bias on  
 event topology is also the smallest in the CDF analysis, where the simplest selection requirements are sufficient to achieve charged-lepton 
 misidentification backgrounds $< 0.4$\%. 

 The amount of QCD radiation accompanying the $W$ bosons and the number of 
 pile-up events are both much smaller at the Tevatron than at
  the LHC.
 Small pile-up implies that the resolution on the inferred $p_T$ of the neutrino, $p_T^{\nu}$ is much better at the Tevatron. Both circumstances force 
 the $W$ boson selection to be $p_T^W < 30$~GeV in the ATLAS analysis, which is suboptimal compared to the $p_T^W < 15$~GeV 
 requirement in the CDF and D\O\ analyses. The former admits more misidentification background, more QCD radiation, more 
 corrections for efficiency-related biases, and considerable smearing of the kinematic distributions sensitive to $M_W$, 
 all of which increase the experimental and theoretical complications associated with
 LHC measurements of $M_W$.  

 One concludes that the CDF experiment requires the smallest corrections for trigger- and selection-induced biases among the hadron
 collider experiments. This is an intrinsic advantage in CDF's favor in terms of robustness of the analysis.  
\subsection{Hadronic recoil measurement}
\label{sec:recoil}
\hspace*{0.18in}
 The $\vec{p}_T^{~\nu}$ vector is inferred by measuring the $\vec{p}_T$ 
 vector of all detectable
 particles, and imposing $\vec{p}_T$-balance assuming a single neutrino in the event. In addition to the charged lepton(s) from $W(Z)$ boson decay, 
 the initial-state QCD radiation which fragments into a spray of hadrons must also be measured (see Fig.~\ref{fig:wzprod}). The  vector sum $\vec{u}$ 
 of these hadrons' $\vec{p}_T$ is referred to as the hadronic recoil vector, since it represents the measurement of $-\vec{p}_T^{~W (Z)}$ using 
 the hadrons. The response and resolution functions that characterize the $\vec{u}$ measurement are more complicated than those for 
 electrons and muons; fortunately the 
  required accuracy is in parts per thousand rather than tens of parts per million for the charged leptons. 

 At a hadron collider, most of the beam particles' energy is transferred to collision fragments that  propagate at small
 angles along the beam pipe and cannot be detected. Hence energy and momentum conservation cannot be used to infer the 
 neutrino's longitudinal momentum; one can work with its $\vec{p}_T$ vector only. 

 Three kinematic quantities are sensitive to $M_W$; the distributions of 
  $p_T$ of the charged lepton and the neutrino, and the transverse mass $m_T$ which is akin to the invariant mass but  uses 
 the leptons' $\vec{p}_T$ vectors. The maximum value of the leptons' $p_T$ (i.e. the distributions' approximate endpoint) 
 is linear in $M_W$, but is also linear in boson $p_T$. 
 The latter is incorporated from an accurate calculation, as discussed in~Sec.~\ref{sec:qcd}, and measured 
 as $\vec{u}$. Both inputs are important because neither is sufficient. 
The special property of the $m_T$ distribution is that its endpoint does not
  depend on the boson's boost, while still being linear in $M_W$. This linear dependence on $M_W$ is used to infer the latter from 
 the shape of these distributions. 

 The $M_W$ extraction from the $p_T^{\nu}$ distribution is by far the most sensitive to the imperfections of $\vec{u}$. 
 Contrary to popular misconception, the use of the charged-lepton $p_T^{\ell}$ distribution 
 is far from immune to the latter. The $m_T$-based extraction 
 has yet another set of correlated dependencies on the various inputs. One of the strengths of the Tevatron analyses is that the 
 benefits of the $m_T$ variable are fully exploited, because the small pile-up maintains good resolution on $\vec{u}$. In 
 fact, as shown in the CDF and D\O\ analyses, the $\vec{u}$-resolution is good enough to permit a reasonably precise extraction even from 
 the $p_T^{\nu}$ distribution~\cite{CDF2022,dzero5fbPRD}.  

 The $\vec{p}_T^{~\ell}$ vector plays the same role in all three kinematic distributions; they differ only in the usage of $\vec{u}$.  
 Hence, the consistency of the $M_W$ values inferred from these three distributions is the most powerful 
 cross-check on the validity of the $\vec{u}$ measurement and the associated uncertainties. The covariances of these $M_W$ values 
 with respect to 
 the response and resolution functions of $\vec{u}$ are used to quantify their mutual consistency. 
 These consistency checks have been published by CDF~\cite{CDF2022,CDF2firstPRD,cdf2fbPRD}. An additional check to examine the consistency
 of $M_W$ measurements in bins of pile-up would be useful for CDF to investigate. 

 Unfortunately, the $\vec{u}$ resolution is quite 
 degraded at the LHC due to the larger pile-up, to the extent 
 that the $m_T$- and $p_T^{\nu}$-based extractions are almost powerless in
 providing a validation of the $\vec{u}$ measurement. Yet, the $p_T^{\ell}$-based extraction, which is essentially the sole
 method of $M_W$ inference at the LHC, is sensitive to a mismeasurement of $\vec{u}$ through the event-selection criteria. 
 As such, the LHC measurements
 of $M_W$ lack the ability to demonstrate their robustness with respect to the hadronic recoil measurement in the same way that CDF
 has demonstrated. The alternatives pursued in the ATLAS analysis are not as powerful~\cite{atlasWmass,atlasWmassErratum}. For 
 example, the $W$-like reconstruction of the $Z$ boson mass (Fig.~16 of~\cite{atlasWmass}) is inconclusive at best. 

 This weakness can be rectified by analysis of the low pile-up data recorded at the LHC for the
 $M_W$ measurement. In the CDF analysis, the $m_T$:$p_T^{\ell}$:$p_T^{\nu}$-based 
 extractions  contribute in the proportion 64:26:10 in the combined final answer, 
 while in the ATLAS analysis the $p_T^{\ell}$-based extraction
  provides 95\% of the final answer~\cite{atlasWmassReanalysis}. Even the $p_T^{\nu}$-based extraction at CDF is more impactful than the $m_T$-based 
 extraction at ATLAS. These weights illustrate the usefulness of the low pile-up data at the LHC. 

 On a related note, the component of $\vec{u}$ parallel to the charged-lepton direction is important; it contributes linearly to the
 leptons' $p_T$. In the CDF analysis, special attention is devoted to the understanding of this component and its
 implications, as documented extensively in each of the three CDF publications from the 
 Tevatron Run 2 data. A corresponding exposition in future ATLAS publications would be welcome, as there is currently 
 little discussion of this key issue.

 The point to emphasize 
 is that the lower pile-up confers a significant advantage to the Tevatron over the LHC, which CDF has fully exploited 
 to show cross-checks of the hadronic recoil measurement. 
\subsection{Likelihood fitting}
\label{sec:likelihood}
\hspace*{0.18in}
Inference of parameters of interest (POI) from data is the domain of statistical analysis. 
 The accurate description of the data is also controlled by many other ``nuisance'' parameters; so named because they 
 need to be constrained in order to extract  the POI, but are  not the motivation for the analysis. When the POI is $M_W$, the nuisance
 parameters  are related to the boson production and decay physics, 
 the detector response and resolution functions and the non-$W$ processes contaminating the data sample. The typical method of 
 statistical inference is likelihood maximization, where the likelihood function depends on the POI(s) and the nuisance parameters. 

 There is a divergence on the choice whether to float the POI and the nuisance parameters separately or simultaneously. 
 Two schools of thought have emerged. The traditional, conservative and time-tested methodology in precision measurements is to float
 them separately. The first step is to 
 constrain all nuisance parameters from independent control samples of data, selected to be strongly constraining
 on the nuisance parameters while being independent of $M_W$. 
 For each nuisance parameter, a causal and explainable logic 
 is developed whose reasoning can be debated and improved, leading to scientific learning
 and progress. Likelihood maximization on these control data with respect to the nuisance parameters provide well-understood constraints
 on the latter. In the next step,  the likelihood with respect to $M_W$ is maximized on the $m_T$ or the lepton $p_T$ distributions
 of the data,  
 while the nuisance parameters are fixed to their respective values determined from causal analysis of the control data. The covariance 
 of $M_W$ in the nuisance parameter space is estimated by varying the 
  nuisance parameters by their {\em a-priori} justified uncertainties and noting the induced shift in the fitted value of $M_W$.  
 The covariance is determined from the model of the data so that it does not suffer from data-driven statistical fluctuations.  
 This methodology, applied by the Tevatron experiments, has steadily advanced our fundamental 
 understanding of the nuisance parameters and their impact on the $M_W$ extraction, as documented in 
 a set of detailed publications~\cite{CDFrun0PRD,CDFrun1aPRD,CDFrun1bPRD,DZEROrun1aPRD,DZEROrun1bCCPRD,DZEROrun1bECPRD,DZEROrun1cPRD,CDF2firstPRD,cdf2fbPRD,dzero5fbPRD,CDF2022}. It is 
 considered a robust, interpretable and transparent method of data analysis because it clearly demarcates between the POI
 and the nuisance parameters and maintains their meaning. 

 The newer school of thought approaches precision analysis as a machine learning exercise. The {\em a-priori} constraints on nuisance 
 parameters from control data are sometimes weak. In an attempt to improve these constraints, a multi-parameter likelihood function
 is constructed that incorporates the nuisance parameters and the POI on an almost equal footing. In this high-dimensional parameter
 space, the likelihood is maximized on 
 the $m_T$ or $p_T^{\ell}$ distributions of the $W$ boson data.  
 In the ATLAS reanalysis of $M_W$~\cite{atlasWmassReanalysis}, the number of nuisance parameters 
 is initially in the hundreds. 

 This method appears sophisticated and increases the accuracy with which the nuisance and the interesting parameters
 together describe the $m_T$ and $p_T^{\ell}$ distributions. However, the interpretation of the inferred 
 $M_W$ and its uncertainties is now opaque since it is mixed with hundreds of other parameters. The causal interpretation of the 
 nuisance parameters may be lost if all parameters including $M_W$ are reduced to mere smoothing knobs 
 to fit a set of kinematic distributions. This raises the concern that $M_W$ is getting tuned to smooth out internal inconsistencies, 
 thereby losing its rigorous interpretation as a measured quantity. Similar concerns arise in the context of ``over-fitting'' in the 
 parametric modeling of data~\cite{overfitting,dyson}. 

 These issues worsen as the dimensionality of the nuisance parameter space increases; even if the initial choice of hundreds of parameters is 
 pruned to 218, as in the ATLAS analysis, the post-fit likelihood function is still impossible to visualize and scrutinize. 
 Mathematically, there is no guarantee that a 218-dimensional likelihood function is well-behaved in the entire domain bounded by the 
 {\em a-priori} uncertainties on the nuisance parameters; i.e. that there is a stable global minimum without the existence of local minima, 
 under-constrained ``valleys'' and saddle points which can lead to chaotic paths in the numerical minimization. Such issues may lead to
 a ``random walk'' of $M_W$ of $\cal{O}$$(\sqrt {N} \delta)$ amounting to tens of MeV, where $N$ is the (large) number of nuisance 
 parameters floated simultaneously in the likelihood function and $\delta$ is a typical $M_W$ deviation of a few MeV induced 
 by each of them.  Is one able to {\it a priori} defend the hypothesis-testing ability of a method which
  relies on a single POI that is susceptible to a random walk in a huge, 200+ dimensional  parameter space? 

The point is that, if the nuisance parameters are not correlated 
with the POI, there is no need to float them simultaneously with the POI. If they are correlated with the 
POI, then it is dangerous to float them simultaneously with the POI. In either case, the more robust 
procedure is to determine nuisance parameters from control samples of data, not the data 
used to determine the POI. 

 The purpose of high-dimensional machine-learning models is to provide a  reproduction of training data, i.e. a method
 of mimicking data rather than a method of learning fundamental principles governing the data. There is a rapidly growing forum in 
 computer science related to interpretability and explainability of black box machine learning models. Explainable models allow human 
 users to comprehend how the model is extracting information from the data and to debate if the model and its results 
 can be trusted. Explainability is a requirement when a parameter that has a fundamental meaning is inferred from the data. 
 CDF has maintained this philosophy of explainability for all its $M_W$ measurements and will continue to do so in the future. 

 In the $M_W$ context, 
 it is important to understand 
 the causal source of information in multi-parameter fitting, such as the reduction of the $M_W$ uncertainty due to proton structure 
 in the ATLAS analysis  from {\em a-priori} 28~MeV 
 to 8~MeV after the multi-parameter fit~\cite{atlasWmassReanalysis}. Which aspects of the $m_T$ and $p_T^{\ell}$ distributions 
  are constraining which degrees of freedom pertaining to the proton structure? Are these constrained 
 degrees of freedom consistent with other world data describing the proton structure? 

 It would be interesting to understand the stability of multi-parameter fitting and to compare what one 
 seems to learn from such fitting to what one expects to learn based on first-principles reasoning. The Fisher
 information is a measure of how much information about any model parameter is carried by the data. Evaluating the Fisher information
 for the nuisance parameters can help understand if multi-parameter fitting is resulting in legitimate or spurious constraints. 

 Using simulated data in which the ground truth is 
 known by design, the likelihood-fitting approaches used by the Tevatron and LHC experiments could be compared. Among other insights, this study 
  could help understand why  re-fitting the same data published in 2018~\cite{atlasWmass,atlasWmassErratum} by ATLAS with identical calibrations 
 caused a 16~MeV shift~\cite{atlasWmassReanalysis} in the extracted $M_W$ -- the only change 
  was that nuisance parameters were floated during 
 $M_W$ extraction. Similarly, $M_W$ changes by 39~MeV in the LHCb analysis when an additional parameter ascribed to the $p_T^W$ spectrum
 is floated simultaneously~\cite{lhcb}. Could these shifts be a symptom of fitting instability in the black-box approach? The spectre of 
 jumping between local minima in the loss function haunts the use of multi-parameter fitting. 
\subsection{Comparisons of $M_W$ analyses at hadron colliders}
\hspace*{0.18in}
The preceding sections discussed key aspects of $M_W$ analyses and comparisons between how different experiments addressed them. 
Table~\ref{tab:comparisons} summarizes these comparisons, along with references to the respective sections. 
\label{sec:comparisonTable}
\begin{table}[!htb]
\begin{center}
\begin{tabular}{|l|cccc|}
\hline 
\hspace{2cm} Criterion   &  CDF & D\O\ & ATLAS & LHCb  \\ 
\hline 
sensitivity to proton structure (Sec.~\ref{sec:qcd}) & low & low & high & high  \\
sensitivity to heavy quark effects (Sec.~\ref{sec:qcd}) & low & low & high & high  \\
electron channel analyzed  & yes & yes & yes & no  \\
muon channel analyzed  & yes & no & yes & yes  \\
check of $p_T^Z$ model with data (Sec.~\ref{sec:qcd}) & yes & yes & yes & yes  \\
check of $p_T^W/p_T^Z$ ratio with data (Sec.~\ref{sec:qcd}) & strong & no & weak & no  \\
emphasis on tracker alignment (Sec.~\ref{sec:pcalib})    & yes & no & no & no  \\
momentum calibration using $J/\psi$ and $ \Upsilon$ (Sec.~\ref{sec:pcalib})      &  yes & no & no & partial  \\
independent $M_Z$ measurement \& confirmation  (Sec.~\ref{sec:pcalib})    &  yes & no & no & no  \\
event selection bias corrections  (Sec.~\ref{sec:selection})    &  small & large & large &  large \\
recoil validation with $m_T$, $p_T^{\ell}$ \& $p_T^{\nu}$ fits (Sec.~\ref{sec:recoil}) & strong & modest & weak & no  \\
$M_W$-independent nuisance parameter constraints  (Sec.~\ref{sec:likelihood})    &  yes & yes & no & no  \\
dimensionality of $M_W$-inference likelihood (Sec.~\ref{sec:likelihood}) & 1 & 1 & 218 & 8 \\ 
explainability of $M_W$ inference procedure (Sec.~\ref{sec:likelihood})  &  high & high & low & low \\
\hline
\end{tabular}
\caption{Comparisons of $M_W$ analyses performed by various experiments at hadron colliders. }
\label{tab:comparisons}
\end{center}
\end{table}
\section{Experience of $M_W$ measurements}
The prerequisite for higher precision measurements of $M_W$ is the increase of the dataset size. It is also a 
 learning experience because reducing the statistical error is only useful if the uncertainties due to nuisance parameters 
 (i.e. systematic uncertainties) are 
 also reduced, and this requires a deeper and more informed analysis. In other words, achieving precision is as much dependent on 
 climbing a long learning curve as having more data. 
\begin{figure}[htbp]
\begin{center}
\vspace*{-32mm}
\includegraphics[width=17.0cm] {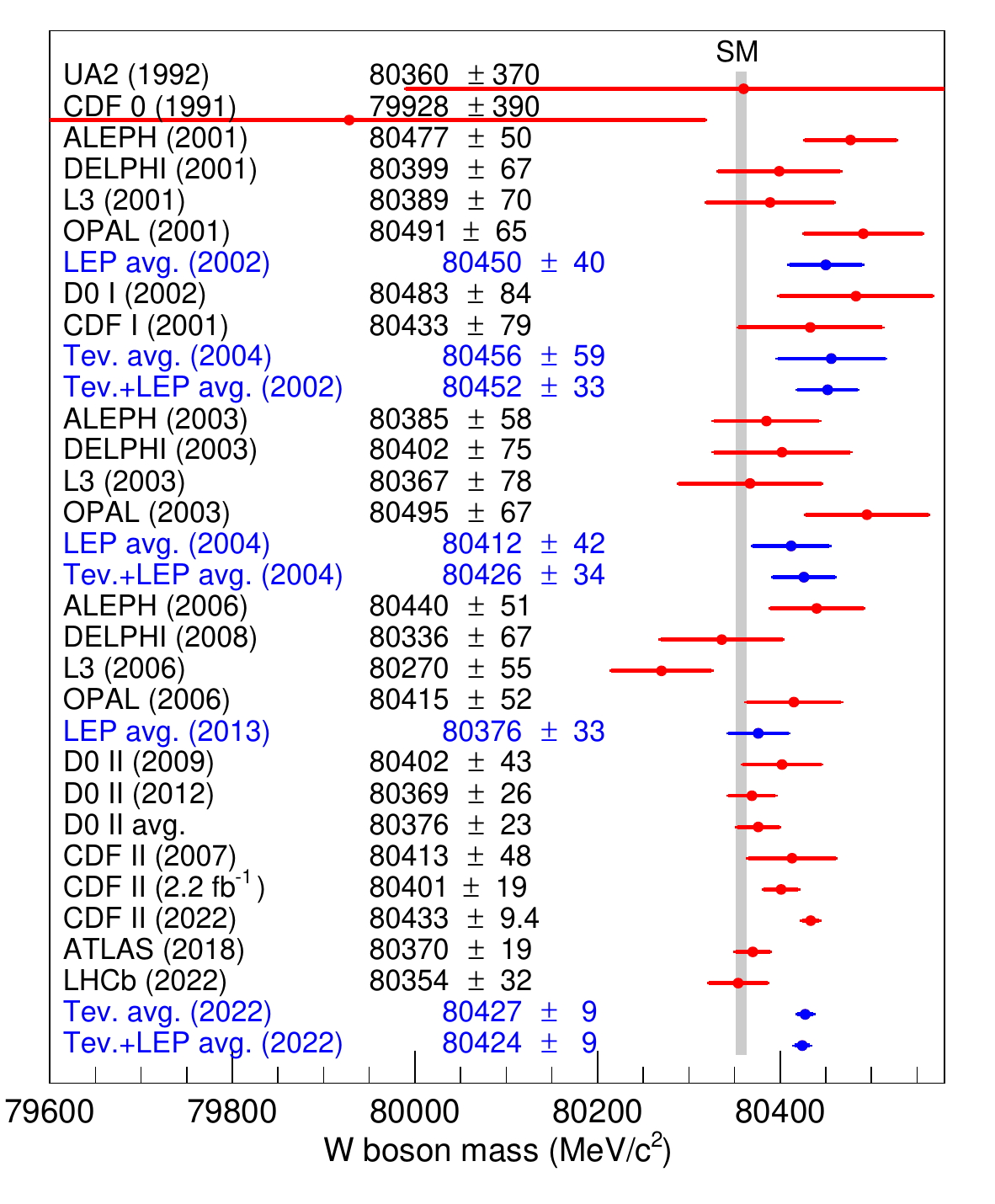}
\end{center}
\vspace*{-14mm}
\caption{{\bf $M_W$ measurements and the SM expectation. }
 The latter includes the estimated uncertainty (4~MeV) due to higher-order effects in perturbation theory, 
 and the uncertainty (4~MeV) induced by the experimental inputs to the calculation, such as the top quark mass. ``Tev.'' refers
 to the CDF and D\O\ experiments at the Fermilab Tevatron. Experimental measurements and their total uncertainties are shown as 
 red points and averages of two or more experimental measurements are shown as blue points. ALEPH, DELPHI, and L3 measurements at CERN LEP are reanalyses of the full dataset (1995-2000), while
 OPAL's 2001 and 2003 results used the first 60\% of the full dataset. D\O\ measurements are published from independent data collected during successive
Tevatron running periods. CDF II measurements are published using cumulative datasets so that all available CDF II data are analyzed for each publication. For the Tevatron experiments, ``I'' and ``II'' refer to the 1992--95 and 2000--2011 running periods respectively. The CDF II (2.2~fb$^{-1}$) value is from~\cite{CDF2022}. The 
 measurements from the UA1 experiment~\cite{ua1Wmass} of $M_W = 82.7 \pm 1.0_{\rm stat} \pm 2.7_{\rm syst} ~ (81.8~^{+6.0}_{-5.3} ({\rm stat}) \pm 2.6_{\rm syst})$~GeV/$c^2$ in the electron (muon) channel would be plotted to the right of this figure. 
}
\label{barPlot}
\end{figure}

The progress on both fronts is notable in Fig.~\ref{barPlot}. 
 Starting with the UA1 and UA2 experiments at the CERN S$p\bar{p}$S collider, 
 hadron colliders have reduced the uncertainties from all sources by two orders of magnitude. During these decades,  better  
 understanding of measurement procedures and nuisance parameters has occasionally shifted the value of a new measurement by 
 more than the estimated uncertainty on the previous measurement. For example, the CDF measurement from the first quarter of the 
 Tevatron Run 2 dataset appears to be shifted low by 33~MeV 
 compared to their high-precision measurement from the full dataset~\cite{CDF2022}. As their analysis
 techniques improved substantially together with a quadrupling of the data, 
 it would be useful for CDF to investigate if there is a causal explanation for the shift, 
 which would contribute to scientific  learning. In the same vein, the shift of 16~MeV in the ATLAS measurement due solely to an 
 updated likelihood estimator, and the shift of 39~MeV in the LHCb measurement due to an additional floated parameter ascribed to the 
 $p_T^W$ spectrum, are worth understanding at a deeper level. 
      
 Measurements of $M_W$ have also been performed at the LEP collider at CERN using two different methods. In the first
 method, $M_W$ is inferred from the rise of the $WW$ 
 pair-production cross section when the electron-positron beam energies are scanned across the $WW$ pair-mass 
 threshold~\cite{ALEPH,DELPHI,L3,OPAL}. 
  These measurements were limited by the dataset size, but this threshold-scan technique is  a robust and accurate method of measuring
 $M_W$. It could be repeated with extreme precision ($\approx 0.5$~MeV) at a future 
 electron-positron collider designed to run at much higher luminosity~\cite{azzuriFCC}. The threshold-scan measurement of $M_W$ is a prime
 motivator for such a future collider. 

 The second method of measuring $M_W$ at LEP used the reconstructed 4-momenta of the final-state hadrons and 
 charged leptons emanating from the decay of the pair-produced $W$ bosons. When both $W$ bosons decay hadronically, strong 
 interactions between the decay products from the two $W$ bosons create a bias in the inferred $M_W$, which is 
 corrected on the basis of hadronization  models~\cite{ALEPH,DELPHI,L3,OPAL}. 
 Methods of attributing the measured particles' momenta
 to the respective $W$ bosons evolved, as did the hadronization models~\cite{alephWmass2000,delphiWmass2001,l3Wmass1999,opalWmass2001}.
   Between 2003 and 2006-8, all four LEP experiments updated their 
 measurements by $\approx 75$~MeV or $\gtrsim 1 \sigma$ each, as seen in Fig.~\ref{barPlot}~\cite{lepEWWG2002,lepEWWG2004,lepwmass}. 
 In preparation for a new electron-positron collider, 
 revisiting the origin of these rather substantial shifts would be educational. 

 The lesson learnt is that understanding and control of nuisance parameters is crucial in  
 all $M_W$ analyses, both at lepton and hadron colliders.  
\subsection{Prospects}
\label{sec:prospects}
The experiments running at the LHC are technologically advanced and can be calibrated with careful analysis of {\em in-situ} 
 control samples of data. They have already collected two orders of magnitude larger data samples than the Tevatron, and
 another order of magnitude increase is expected from future running. However, these samples will suffer from a large rate
 of pile-up events because they have been or 
 will be collected with a high beam luminosity which enables the large signal sample rate. Ironically,
 the requirements of the $M_W$ measurement from the perspective of precision and robustness are better served by a lower beam energy
 to simplify the $W$ boson production dynamics and a lower beam luminosity to reduce pile-up. While the technical and scientific cost
 to the rest of the LHC physics program may outweigh the benefit of a lower beam energy, a short-term reduction in the beam 
 luminosity is an attractive option. Data collected at low luminosity will reduce 
 the number of additional proton-proton collisions that complicate the inference of the neutrino momentum. Among other benefits, 
 low-luminosity data will provide the built-in consistency check of $M_W$ extracted from the $m_T$, $p_T^{\ell}$ and 
 $p_T^{\nu}$ distributions, which is only possible when the $p_T^{\nu}$ resolution is not degraded by pile-up. 

 In the longer
 term, an electron-positron collider that serves as a Higgs and electroweak factory will be the ultimate precision machine
 for the measurement of $W$, $Z$ and Higgs boson properties~\cite{ilc,cepc,clic,blondelFCCee}. The beam-energy scan at the $WW$ pair-production threshold is the most robust method 
 of confirming the difference between the measured $M_W$ and its SM-calculated value~\cite{azzuriFCC}. This difference is most  striking in the recent  
 CDF measurement, but Fig.~\ref{barPlot} shows the historical trend for this difference to be positive.  

 The $M_W$ measurement at this future collider complements an extensive program of precision measurements of $Z$ bosons which can 
 improve upon the precision achieved at LEP by a factor of 20--100~\cite{blondelFCCee}. Similarly, certain properties of the Higgs boson can be measured
 with precision ranging from 1\% to 0.2\%, exceeding the capability of the LHC in these  respects~\cite{blondelFCCee}. 
 As with $M_W$, the  motivation for making extremely precise
 measurements of such observables is that they can be calculated with commensurately high precision in the SM. 
 A pattern of deviations between measured and calculated values may emerge that guides us to develop a 
 deeper theory that subsumes and supersedes the SM.  
\subsection{Weak mixing angle}
In Sec.~\ref{sec:motivation} we discussed how the SM predicts the value of $M_W$ when other key parameters are fixed by measurements. 
 We mentioned the weak mixing angle that describes the superposition of two generators of the SU(2)$_{\rm weak}$ and 
 U(1)$_{\rm hypercharge}$ gauge groups respectively. The SM also predicts the weak mixing angle when the above-mentioned 
 parameters are fixed~\cite{sirlin}; thus, a precise measurement of 
 the weak mixing angle provides a stringent test of the SM, along the same lines as and independent of $M_W$. 

\begin{figure}[htbp]
\begin{center}
\vspace*{-15mm}
\includegraphics[width=17cm]{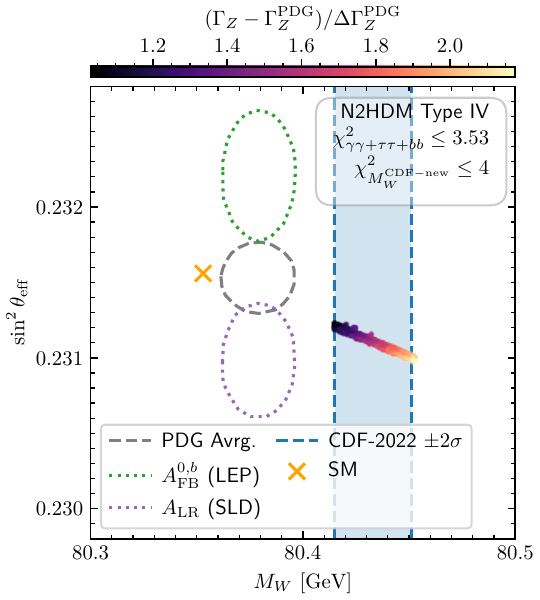}
\caption{Illustration (reproduced with permission from Fig.~2 of Ref.~\cite{sven}) of the two most precise measurements of the weak mixing angle $\theta_{\rm eff}$ (shown conventionally as $\sin^2 \theta_{\rm eff}$ on the vertical axis) from the LEP experiments and the SLD experiment respectively~\cite{weakMixingAngle}, and their average~\cite{pdg}.  
 The CDF measurement of the $W$ boson mass is shown on the horizontal axis. The SM expectation is shown as the orange cross. 
 The SLD measurement of $\sin^2 \theta_{\rm eff}$ and the CDF measurement of $M_W$ concur on a set of new-physics scenarios shown as 
 the line of colored points. While a supersymmetric framework with 
 additional Higgs bosons is used in~\cite{sven}, the common feature
 between these and other models is the violation of $WZ$ custodial symmetry mentioned in Sec.~\ref{sec:motivation}. }
\label{fig:angle}
\end{center}
\end{figure}

 The two most precise measurements of the weak mixing angle have been performed in the 1990s by the LEP experiments and by the 
 SLD experiment at the SLAC SLC accelerator respectively~\cite{weakMixingAngle}. The measurements were based on different properties of $Z$ bosons; the SLD observable depends on leptonic couplings only, while the LEP observable involves $b$-quark  
 couplings. The measurements differ by $3.2 \sigma$. The SLD measurement relies on beam polarization and the LEP measurement relies 
 on identification of $b$-quarks produced from the $Z$ boson decay. 

 This discrepancy has not been resolved; nevertheless the two measurements have been averaged.  Additional measurements have been 
 made in atomic parity violation, electron and neutrino scattering, and at the Tevatron and LHC (see~\cite{pdg} for a review).  
 Analysis of the available LHC data may yield a precision similar to that of the SLD measurement.
  New experiments~\cite{moller,solid,p2} including those at a future electron-positron collider~\cite{ilc,cepc,clic,blondelFCCee} plan to 
 improve upon the precision of this parameter. 

 In the meantime, the status of this discrepancy in the weak mixing angle is similar to the $\sim 3 \sigma$ difference between the CDF and ATLAS measurements of 
 $M_W$.  Per precedence, the scientific approach is to combine the $M_W$ measurements using established statistical methods while 
 continuing investigations. This approach has been applied when even larger differences, such as the $5.5 \sigma$ 
 between measurements of the fine structure constant $\alpha$ using Cesium and Rubidium atoms, have been found~\cite{pdg}. An 
 alternate, biased approach has been presented in~\cite{mWcombination}. Unless scientific concerns on any published result are documented, 
 it is not
scientific practice to isolate any measurement as an outlier and remove it from a world average. 
 The Particle Data Group has maintained a time-tested procedure for
combining measurements that are not in perfect agreement - the PDG review is replete
with examples~\cite{pdg}.

 It is interesting to note that the SLD measurement of the weak mixing angle and the CDF measurement of $M_W$ each point towards the 
 same new physics, qualitatively and quantitatively. This concurrence (Fig.~\ref{fig:angle}) has been pointed out 
 in~\cite{sven}, where it is shown that both measurements favor a similar violation of the $WZ$ custodial symmetry that we discussed in
 Sec.~\ref{sec:motivation} as an approximate feature of the SM. 
\subsection{Muon magnetic moment}
Analogous to $M_W$ and $\sin^2 \theta_{\rm eff}$, the magnetic moments of the electrons and the muon are precisely measurable  
 and their values are predicted by the SM. 
 The magnetic moment is proportional to the intrinsic spin, the charge/mass ratio and the $g$-factor, where $g=2$ in Dirac's theory of the electron. 
 In quantum field theory this prediction is modified by quantum fluctuations which are 
 calculable as a perturbation series with the electromagnetic fine structure constant $\alpha$ as the expansion parameter. 
 A measurement of $(g-2)$ is a stringent test of the SM and has been pursued for both electrons and muons for many decades, along with 
 increasing accuracy of its calculations. 

 In recent years, a difference between the measured and calculated values of muon $(g-2)$ 
 has been reported~\cite{fnalMuPRLLatest,fnalMuPRL,fnalMuPRD,bnlMu}. The most recent measurement from Fermilab~\cite{fnalMuPRLLatest} 
 improves the precision of the experimental world average by a factor of two. Simultaneously, the SM calculation of this quantity is being 
 revisited~\cite{castelvecchi} due to improved inputs from QCD calculations on the lattice~\cite{bmwLattice,lehner} and from a new measurement of the
 $e^+ e^- \to \pi^+ \pi^-$ cross section near threshold~\cite{cmd3}. This information changes the hadronic vacuum polarization and the 
 resulting update to $\alpha$ would bring the SM value of the muon $(g-2)$ closer to 
 the measured value. 
  
 The effect of low-energy hadronic quantum fluctuations on $\alpha$ propagate to $M_W$ and to the muon $(g-2)$ in opposite  
 directions~\cite{gm2Mw}. In other words, if the discrepancy between the measured and calculated
 values were to reduce for the muon $(g-2)$ due to an adjustment of $\alpha$, the discrepancy must increase for {\it all} $M_W$ 
 measurements, and vice versa. 
 There is no way to reconcile both discrepancies simultaneously by adjusting this hadronic quantum correction on $\alpha$. 

The interesting conclusion is that the combination of these two measurements makes a striking case for new
physics. Whether the new physics manifests in
the muon's magnetic moment, or the $W$ boson mass, or both, these two measurements are hinting at 
a future beyond the Higgs boson. 
\section{Summary}
Developed over a century, the SM is one of the crowning achievements of science. It is built on a remarkable 
 set of axioms of local quantum field theory, special relativity and group theory, and has been both strongly guided by and predictive
 of a vast set of experimental observations. The SM's theoretical structure and past consistency with experimental data together 
 led to the prediction of the Higgs boson, which was spectacularly confirmed by its observation at the LHC in 2012. 
 
 Nevertheless, the SM is not expected to be the most fundamental theory of matter and forces. The Higgs sector of the SM is not as 
 axiomatic as the rest of the theory; the Higgs field's fermionic- and self-interactions 
 are parametrized phenomenologically without a fully
 dynamical construction. Second, the parameter describing the Higgs boson mass receives large quantum corrections
 that are very sensitive to unknown physics at arbitrarily high energy scales.  The small observed value of the Higgs boson mass   
 raises the concern of  
 ``naturalness'' of the Higgs sector, because it requires an extremely tuned set of conditions at higher energies. Such ``fine-tuning'' 
 is prevented in other sectors of the SM by its symmetry structure~\cite{symmetryNaturalness}, and can be alleviated for the Higgs 
 mass parameter as well by extending the field content and symmetries of the theory. The fine-tuning associated with the 
 Higgs mass increases quadratically with the energy scale of new physics, which creates an expectation that the new physics is around 
 the corner. 

 Another set of deficiencies in the SM relate to 
 crucial facts of cosmological significance - the observed excess of matter over antimatter in the universe, dark matter, 
 dark energy and cosmic inflation, all of which are beyond the scope of the SM. Finally, the force of gravity cannot be accommodated 
 into the SM's quantum field-theoretic framework. 

 For such reasons, physics beyond the SM is expected to show up. {\it A-priori} there is no definitive prediction for the energy scale
 at which new physics may be revealed. Nevertheless,  
 these considerations justify current and future particle physics experiments that are designed to observe phenomena 
 not explainable by the SM. 
  
 We discussed the theoretical reasoning for the $W$ boson mass to be a sensitive probe of certain aspects
 of new physics, which explains why this measurement has been pursued with increasing precision (progressing 
 by two orders of magnitude) over the 
 last three decades. Finally we have an $M_W$ measurement that shows a  significant tension; it
  is the largest deviation observed from a prediction of the SM. The most exciting possibility is that the new 
 physics that this deviation points 
 towards will reveal itself as new particles in the near future. 

\vspace{3mm}

\textbf{Acknowledgements}

\vspace{1mm}

We thank  Kaustubh Agashe, Sunil Gupta and Ronen Plesser for helpful discussions.

\end{document}